\documentclass{ptephy}

\usepackage[dvips]{pict2e}
\usepackage{booktabs}  

\begin{document}

\title{Possible Existence of Charmonium-Nucleus Bound States}

\author{\name{Akira Yokota}{1, \ast}, \name{Emiko Hiyama}{2}, 
and \name{Makoto Oka}{1}}

\address{
\affil{1}{Department of Physics, Tokyo Institute of Technology, 2-12-1, Ookayama, 
Meguro 152-8551, Japan} 
\affil{2}{RIKEN Nishina Center, 2-1, Hirosawa, Wako, Saitama 351-0198, Japan} 
\email{yokota.a.aa@m.titech.ac.jp}
}

\begin{abstract}
Possible existence of $(c\bar{c})-$nucleus bound states are examined.
We adopt Gaussian potentials for the $\eta_{c}-N$ and $J/\psi-N$ interactions. 
The relations between the scattering lengths $a$ of $(c\bar{c})-N$ interactions 
and the binding energies of 
$\eta_{c}-NN$, $J/\psi-NN$ and $J/\psi-^{4}$He are given. 
The results show that scattering lengths $a \le -0.95$ fm are needed to make 
$\eta_{c}-NN$ and $J/\psi-NN$ bound states, 
while for $a \le -0.24$ fm there may exist a $J/\psi-^{4}$He bound state.
\end{abstract}


\maketitle

\section{Introduction}
Recent studies of hadronic interactions at low energy have revealed that 
QCD allows wide variety of bound and/or resonance states of hadrons.
For instance, a strong attraction in the S-wave $\bar K -N$ system, which is derived from the chiral symmetric interaction of 
the kaon as a Nambu-Goldstone boson, seems to form the $\Lambda(1405)$ baryon resonance in its $I=0$ channel \cite{Hyodo2012}.
The same interaction will further generate $\bar K-$nucleus bound states, which are the recent subject of intense study.
Similar bound states may appear in systems with heavy quark mesons, 
such as the $D-N$ and $B-N$ systems \cite{Bayar2012}. 
Furthermore, some of the newly observed quarkonium-like states are found to be better described as bound or resonance states of 
two heavy-quark mesons \cite{Godfrey2008}.
Here a naive expectation is that due to the large masses of the heavy hadrons, more bound/resonance states 
may exist in the heavy-quark sector.

Such hadronic bound states are very useful in studying the low-energy hadronic interaction which cannot be accessed directly from experiment.
For instance, in hypernuclear physics, various interactions of hyperons, where two-body scattering experiments are not available, 
are determined by the spectroscopy data of light and heavy hypernuclei \cite{Hashimoto2006}.
Precise knowledge of those interactions is important as they play crucial roles in understanding properties of dense and 
hot hadronic matter produced in the early universe or heavy ion collisions as well as compact stars.

In this context, heavy quarkonia, $c\bar c$ or $b\bar b$ states, may have some new aspects and advantages.
Let us consider the ground-state charmonium ($c\bar{c}$) states, 
$J/\psi$ $(J^{\pi}=1^{-})$ or $\eta_c$ $(0^{-})$.
Their interactions with the nucleon $N$ 
are quite different from the other hadronic interactions. 
First, the charmonium and the nucleon 
have no valence quarks in common, 
so the interactions mediated by flavor singlet meson exchanges are strongly suppressed by the OZI rule.
For the same reason, the Pauli exclusion principle for quarks does not cause 
a repulsion at short distances.
Further, a single gluon exchange is prohibited since $(c\bar{c})$ and $N$ are both color singlet.
Thus the $(c\bar{c})-N$ interactions are dominated by multi-gluon exchanges.
The QCD color van der Waals interaction, which is a typical multi-gluon exchange interaction, was applied to $(c\bar{c})-N$ system 
by Brodsky\cite{Brodsky1990}.
Studies by Luke et al.\cite{Luke1992}, Kharzeev\cite{Kharzeev1994} 
and Brodsky\cite{Brodsky1997} showed 
that the $(c\bar{c})-N$ interaction is attractive.

Recent lattice QCD calculation\cite{Kawanai2010} showed that 
the $(c\bar{c})-$nucleon scattering lengths correspond to weak attractive interactions
and thus confirmed the previous studies.
It is therefore interesting to see whether such attraction is enough to form bound states of $(c\bar c)$ and nucleus.
Indeed, considering the heavy mass of the $(c\bar{c})$, 
it is highly possible that $(c\bar c)-$nucleus bound states exist, 
when the nucleon number $A$ increases\cite{Brodsky1990, Wasson1991}.
The situation may be similar to the case of $\Lambda$ hypernuclei. 
There is no two-body $\Lambda N$ bound state, 
while three or multi-baryon systems with $\Lambda$ have bound states. 
It is found that the spectroscopy of such hypernuclei provides us with the strengths and detail structures of the $\Lambda - N$ interaction.
Replacing $\Lambda$ by $(c\bar{c})$ or $J/\psi$ may also reveal 
the dynamics between the charmonium and the nucleon, the ordinary matter. 
For instance, it is interesting to see how strong is the spin-dependent force involving charm quarks. 
It is naturally considered that the magnetic gluon coupling is suppressed by the $1/m_{Q}$ factor, 
and the spin dependent force of charmonium is weak. 
With the spin-one $J/\psi$ bound in nuclei, we will be able to determine the spin dependent forces, 
i.e., the spin-spin, spin-orbit and tensor components, of $J/\psi$ and $N$, quantitatively.
We can also see how the $\eta_{c} - $nucleus states mix with the $J/\psi-$nucleus.

In this paper, we consider bound states
of the charmonium with few-nucleon systems. 
A standard approach would take a microscopic model or calculation of the 
$(c\bar{c})-N$ interaction and apply it to a nuclear system.
It may, however, be not conclusive because the model settings and parameters 
have ambiguities and their relations to the obtained binding energies are complex.
We instead employ a phenomenological potential model to represent the charmonium-nucleon interaction and derive relations between the $(c\bar{c})-N$ scattering length  
and the potential parameters. 
We then apply the potential to a nuclear system 
and see whether such a potential gives a bound state or not.
Concretely, we consider the $(c\bar{c})-NN$ and $(c\bar{c})-^{4}$He cases 
and apply the Gaussian Expansion Method (GEM) 
\cite{Kamimura1988, Hiyama2003} to obtain their binding energies and the wave functions. 
Then we determine the values of the scattering length which is needed 
to form a bound state of $(c\bar{c})$ with the deuteron or $^{4}$He.
In the end, we find that the scattering length given by the recent lattice QCD
may be large enough to make a bound state of $J/\psi$ in $^{4}$He.

In section 2,
we explain our strategy to relate the scattering length of $(c\bar{c})-$nucleon system 
to the phenomenological potential that we adopt in the nuclear calculation. 
Then we give a formulation for calculating $(c\bar{c})-$nucleus bound states in section 3.
We introduce Gaussian potentials as effective $(c\bar{c})-N$ interactions.
In section 4, we show the calculation results. 
The relations between the scattering lengths and the potential strengths 
for $\eta_{c}-N$ and $J/\psi -N$ are given first, 
and subsequently the relations between the scattering lengths and the binding energies of 
$(c\bar{c})-$nucleus are obtained for the deuteron and $^{4}$He.
In section 5, 
we discuss $J/\psi - \eta_{c}$ mixing and the decay of charmonium in nuclei. 
In section 6, 
summary and conclusions are given. 


\section{$(c\bar{c})-N$ interaction}
In the previous studies,  
Brodsky~\cite{Brodsky1990}, Wasson~\cite{Wasson1991} and Belyaev~\cite{Belyaev2006}
assumed a Yukawa-type $c\bar{c}-N$ potential
\begin{equation}
  v ( r ) = A \ \frac{e^{-\alpha r } } { r } \ ,
  \label{eq:YukawaPotential}
\end{equation}
while de T\'eramond \cite{Teramond1998} additionally assumed a Gaussian type potential
\begin{equation}
  v ( r ) = v_{\rm eff} \ e^{-\mu r^{2}}   \label{eq:GaussPotential},
\end{equation}
where $A$ and $\alpha$ in the Yukawa type potential, 
and $v_{\rm eff}$ and $\mu$ in the Gaussian type potential are the parameters
which represent the strengths and the ranges of the potentials, respectively.
Usually, these parameters are taken to fit experimental data or microscopic calculations 
of the scattering length and the other observables of the system. 
Under the situation with only very limited experimental data, 
we need to determine or assume the values of these parameters 
by considering the physical properties of the system
to calculate $c\bar{c}-$nucleus binding energy $B$, 
as the previous researches have done~\cite{Brodsky1990,Wasson1991,Belyaev2006,Teramond1998}.

Here, we take a different strategy.
Since the $(c\bar{c})-N$ interaction is weak and has short range,
the scattering length $a$ represents the interaction well 
and the binding energy $B$ of $(c\bar{c})-$nucleus 
does not depend so much on the details of the form of the potential.
Thus we calculate $B$ as a function of $a$.
The good point of this approach is that 
it gives the correspondence between 
the binding energy $B$ of $(c\bar{c})-$nucleus 
and the $(c\bar{c})-N$ scattering length $a$.
Therefore, whatever the details of the true $(c\bar{c})-N$ potential are,
as far as they are weak and short ranged, the relation obtained will be valid.
We will show in later chapter that the range dependence of binding energy decreases
as the interaction becomes weaker.


In our present approach, 
we employ the Gaussian Expansion Method (GEM)~\cite{Kamimura1988, Hiyama2003} 
for few-body charmonium-nuclear systems.
This is a variational method which gives the upper limit of the energy eigenvalues.
So once we obtain a bound state ($E<0$) for a given Hamiltonian with some trial function, 
then the true binding could only be deeper for the same Hamiltonian.
This is suitable for finding a shallow bound state.
In general, GEM provides very accurate eigenvalues and wave functions of few-body systems.
Then the problem reduces to whether we could give correct Hamiltonian,
the form of the potentials and the values of parameters.

%

From a microscopic point of view,
it is important to calculate the $(c\bar{c})-N$ scattering length 
from the first principle, such as lattice QCD. 
Recently, Kawanai and Sasaki evaluated the scattering lengths of
$\eta _{c}-N$, $J/\psi-N$ (total spin $J=1/2$), and $J/\psi-N$ ($J=3/2$) 
in the quenched lattice QCD \cite{Kawanai2010}.
They found $a^{J/\psi -N}_{\rm SAV} \simeq -0.35$ fm and $a^{\eta _{c}-N} \simeq -0.25$ fm 
with an error of around $0.1$ fm
\footnote{We convert the sign of the scattering lengths to our definition, 
$\displaystyle \lim_{k \to 0} k\cot \delta=-\frac{1}{a}$.}, 
where SAV stands for the spin-averaged value 
$(1/3)(a_{J=1/2}^{J/\psi -N} + 2 a_{J=3/2}^{J/\psi -N})$.
This shows that charmonium-nucleon interaction is weak but attractive.
Although the error bars overlap with each other,
the tendency of
$a^{\eta _{c}-N} \gtrsim a_{J=1/2}^{J/\psi -N} \gtrsim a_{J=3/2}^{J/\psi -N}$
can be seen in the result (see Fig. 5 of \cite{Kawanai2010}).
This implies that the spin-spin interaction is not zero but weak compared to the central force. 
It is interesting to see that 
the spin $S_{J/\psi -N}=3/2$ potential  
has a stronger attraction than $S_{J/\psi -N}=1/2$.


\section{Formulation}
In present analysis, 
we consider $\eta_{c}$ $(J^{\pi}=0^{-})$ and $J/\psi$ $(1^{-})$ for the charmonium, 
both of which are charge neutral and have no electromagnetic interaction.
Since the $(c\bar{c})-N$ interaction due to the QCD color van der Waals interaction 
is weak, states with lower orbital angular momentum $L$, 
especially the s-wave states ($L=0$), are important. 
Also we are searching charmonium-nucleus bound states and their binding energies. 
Thus we only calculate the ground state with $L=0$.
Then, the spin-orbit force does not contribute.
Furthermore, for simplicity, we do not consider the tensor force, 
which is supposed to be weaker than the central force.

\subsection{$(c\bar{c})-N$ potential}

First, we consider an effective $(c\bar{c})-N$ potential.
For $\eta_{c}-N$ potential, 
we only have a spin-independent central force, given by a single Gaussian, as
\begin{alignat}{2}
  v_{\eta _{c}-N} ( r ) &= v_{0}e^{-\mu r^{2}}.  
  \label{eq:Etac_N_potential}
\end{alignat}

For the $J/\psi-N$ potential, 
we introduce not only the spin-independent force but also the spin-spin interaction 
in order to take into account the spin structure of the $J/\psi-N$ interaction,
\begin{equation}
  v_{J/\psi-N} ( r ) = v_{0}e^{-\mu _{1} r^{2}}
                           + v_{s}(\mathbf{S}_{J/\psi} \cdot \mathbf{S}_{N})e^{-\mu _{2} r^{2}}.
\label{eq:Jpsi_N_potential1}
\end{equation}
Here, $\mathbf{S}_{J/\psi}$ and $\mathbf{S}_{N}$ are 
the spin operators of the $J/\psi$ and $N$, respectively.
In principle, the parameters $\mu _{1}$ and $\mu _{2}$, 
which represent the ranges of the spin-independent part and 
spin-spin part of the potential respectively,
could be different from each other.
But here the details of the shape of the potential is not so important, 
so we assume for simplicity $\mu _{1} = \mu _{2}$ ($\equiv \mu$).
This assumption is reasonable because both the spin-independent and -dependent potentials 
come from the gluon exchanges and thus the ranges must be similar.
This is in contrast to the nuclear force, 
where exchanges of various mesons with different masses contribute to the potential.
Even if we choose $\mu_{1} \neq \mu_{2}$, the results in this paper 
do not change qualitatively. 
As we see later in 
Figs. \ref{fig:Range_dependence1}, \ref{fig:potential_form_dependence} and \ref{fig:J_4He_E_a},  
the relations between the $(c\bar{c})-N$ two-body scattering lengths and 
the $(c\bar{c})-$nucleus binding energies 
are insensitive to the change of the potential ranges.

Then, the potentials for the total spin $S_{J/\psi-N}=1/2$ and $3/2$ states 
are given by
\begin{alignat}{5}
v_{J/\psi-N}(r) &=  (v_{0}+v_{s}(\mathbf{S}_{J/\psi} \cdot \mathbf{S}_{N} ))e^{-\mu r^{2}} 
                         \label{eq:Jpsi_N_potential2_1} \\
                     & \equiv v_{\rm eff}(S_{J/\psi-N}) e^{-\mu r^{2}}, 
                        \label{eq:Jpsi_N_potential2}  \\
v_{\rm eff}(S_{J/\psi-N})&= 
   \begin{cases}
        v_{0}-v_{s}    \quad \quad  \bigl(S_{J/\psi-N}=1/2\bigr)           \\
        v_{0}+\frac{1}{2}v_{s}  \quad \ \bigl(S_{J/\psi-N}=3/2\bigr).    \\
   \end{cases}                          
\end{alignat}

The $(c\bar{c})-N$ scattering length is calculated by solving the Schr\"odinger equation, 
\begin{alignat}{2}
\Bigl( -\frac{1}{2\mu} \nabla^{2} + v_{c\bar{c}-N} \Bigr) \psi = E \psi
\end{alignat}
at $E=0$, where we assume the non-relativistic kinematics 
and the reduced mass is 
$\mu = ( M_{N} M_{c\bar{c}} ) /( M_{N} + M_{c\bar{c}} ) = 713.7$ MeV for $\eta_{c}-p$ and 
$720.1$ MeV for $J/\psi-p$.
%
%
\subsection{$(c\bar{c})-NN$ systems}
Using various values of the $(c\bar{c})-N$ potential parameters, 
we calculate the $(c\bar{c})-NN$ three-body system 
using the Gaussian Expansion Method (GEM).
Three sets of Jacobian coordinates for the three-body systems of $(c\bar{c})-NN$ are 
illustrated in Fig.1, in which we further take into account the antisymmetrization between 
two nucleons. 
The total Hamiltonian and the Schr\"odinger equation are given by 
%
\begin{alignat}{2}
&(H-E)\Psi_{JM}=0 \\
&H=T+V_{N_1-N_2}+v_{c\bar{c}-N_1}+v_{c\bar{c}-N_2}
\end{alignat}
%
where $T$ is the kinetic-energy operator and 
$V_{N_1-N_2}, v_{c\bar{c}-N_1}, v_{c\bar{c}-N_2}$ 
are the potentials between $N-N$ and $(c\bar{c})-N$, respectively.
For the $N-N$ interaction, we employ the Minnesota potential 
\cite{Reichstein1970, Thompson1977},
which consists of two Gaussian terms and has only central forces.
Contribution of the tensor force is effectively included in the central potentials.
The potential parameters 
are adjusted to reproduce the binding energy of the deuteron. 
The use of Minnesota potential is adequate  
as far as the $(c \bar{c}) -N$ interaction is weak 
and nucleon density distribution in nucleus (deuteron) 
is not affected by $(c \bar{c})$ significantly. 
This applies to a case in which $(c\bar{c})-NN$ having a shallow bound state.

\begin{figure}[!h]
  \centering
  \begin{picture}(400,105)
    \linethickness{1.5pt}
    \multiput(60,82)(140,0){3}{\circle{15}}
    \multiput(30,30)(140,0){3}{\circle{15}}
    \multiput(90,30)(140,0){3}{\circle{15}}
       \put(60,82){\vector(100,-175){30}}
       \put(75,56){\vector(-300,-175){45}}
       \put(170,30){\vector(100,175){30}}
       \put(185,56){\vector(300,-175){45}}
       \put(370,30){\vector(-10,0){60}}
       \put(340,30){\vector(0,10){52}}
    \multiput(20,10)(140,0){3}{$N_{1}$}
    \multiput(90,10)(140,0){3}{$N_{2}$}
    \multiput(51,97)(140,0){3}{$(c\bar{c})$}
       \put(80,58){$\mathbf{r_{1}}$}
       \put(38,45){$\mathbf{R_{1}}$}
       \put(170,58){$\mathbf{r_{2}}$}
       \put(210,45){$\mathbf{R_{2}}$}
       \put(332,20){$\mathbf{r_{3}}$}
       \put(345,58){$\mathbf{R_{3}}$}
    \put(50,0){$c=1$}
    \put(190,0){$c=2$}
    \put(330,0){$c=3$}
  \end{picture}
  \caption{Three Jacobian coordinates of the three-body system.}
  \label{fig:Jacobi}
\end{figure}
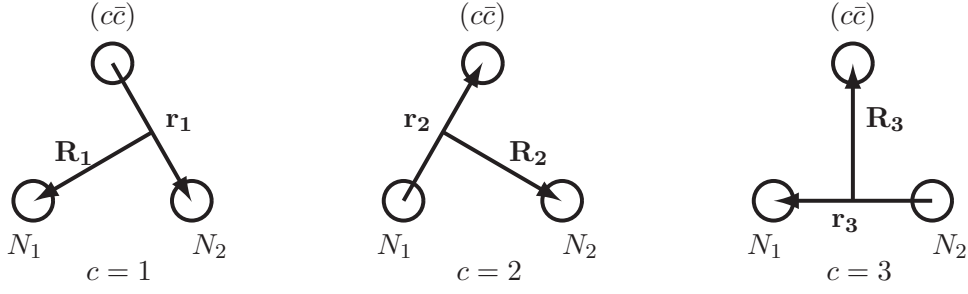

The total wave function is expanded in GEM as 
\begin{equation}
  \Psi_{JM} = \sum_{c=1}^{3} \sum_{n=1}^{n_{max}} \sum_{N=1}^{N_{max} } \sum_{I}
           C_{nNI}^{c} \phi_{nlm}^{c}(\mathbf{r_{c} } ) \psi_{NLM}^{c} (\mathbf{R_{c} } )
           [ [\chi_{s}(1)\chi_{s}(2)]_{I} \chi_{s}(3)]_{JM}
\end{equation}
\begin{alignat}{5}
  \phi_{nlm}^{c}(\mathbf{r}) &= r^{l} e^{-\nu_{n} r^{2} } Y_{l}^{m}(\hat{\mathbf{r}})  \\
  \nu_{n}&= \frac{1}{r_{n}^{2}},  \qquad
  r_{n} = r_{1}a^{n-1}   \quad (n = 1, \dots, n_{max} )   \label{eq:range1}     
\end{alignat}
\begin{alignat}{5}
  \psi_{NLM}^{c}(\mathbf{R}) &= R^{L} e^{-\lambda_{N} R^{2} }  Y_{L}^{M}(\hat{\mathbf{R}})  \\
  \lambda_{N}&= \frac{1}{R_{N}^{2}},     \quad
  R_{N} = R_{1}A^{N-1}   \quad (N = 1, \dots, N_{max} )   \label{eq:Range2}
\end{alignat}
where the Jacobi coordinates $r_{c}$ and $R_{c}$ ($c=1,2,3$) 
are taken as shown in Fig. \ref{fig:Jacobi} 
and $\chi_{s}(1)$, $\chi_{s}(2)$ and $\chi_{s}(3)$ are the spin wave functions of the particles 1, 2 and 3.
The orbital angular momenta $l,m$ and $L,M$ correspond to $r$ and $R$, respectively. 
The number of the basis functions used in the present calculation are 
$n_{max}^{(c)}=10$ and $N_{max}^{(c)}=10$ 
for $c = 1, 2$ and 
$n_{max}^{(c)}=12$ and $N_{max}^{(c)}=14$ for $c = 3$.

%
We calculate 
the s-wave bound state of $\eta_{c}$ and the deuteron (spin $S_{NN}=1$ and isospin $T=0$) 
for the $\eta_{c}-N$ potential Eq. (\ref{eq:Etac_N_potential}).
Then, the only possible state for $\eta _{c} -NN$ has
the total angular momentum $J=1$, $S_{NN}=1$ and $T=0$.
We do not consider the $T=1$ state, as the $NN$ ($S_{NN}=0$, $T=1$) state 
has a weaker attraction.
%

\begin{table}[!h]
    \centering
    \begin{tabular}{cccc} \hline 
     $T$   \quad    $J$     &  $S_{NN}$     &    $S_{J/\psi -N}$     \\
     \hline \addlinespace[2pt]
       0    \quad      0       &      1           &    $1/2$   \\ 
     \addlinespace[2pt]  \hline \addlinespace[3pt]
       0    \quad      1       &      1           &    $1/2, \ 3/2$ \\
     \addlinespace[3pt]  
       1    \quad      1       &      0           &    $1/2, \ 3/2$ \\  
     \addlinespace[3pt] \hline \addlinespace[2pt]  
       0    \quad      2       &      1           &    $3/2$ \\ 
     \addlinespace[2pt] \hline 
    \end{tabular}
    \caption{Possible $(T,J)$ combination of the $J/\psi -NN$ system}
    \label{JpsiNN_spin_states}
\end{table} 

The binding energy of $J/\psi-NN$ system is affected by the spin-dependence 
of the $J/\psi-N$ interaction in Eq. (\ref{eq:Jpsi_N_potential2_1}). 
We consider three channels, 
$J^{\pi}=0^{-}, 1^{-}$ and $2^{-}$ with $T=0$ 
for the three-body system.
In Table \ref{JpsiNN_spin_states}, 
we illustrate the combination of the spins of $NN$ and $J/\psi N$, 
for each $J$ and $T$. 
We do not calculate $T=1$ channel as before.

One sees that for $J^{\pi}=0^{-}$ and $J^{\pi}=2^{-}$, 
$S_{J/\psi-N}$ is uniquely given as $1/2$ and $3/2$, respectively.
On the other hand, for the $J^{\pi}=1^{-}$, $T=0$ state, 
both $S_{J/\psi-N}=1/2$ and $3/2$ are mixed as 
\begin{align}
   \Big|  (NN)_{S_{NN}=1} J/\psi ; J=1 \Big\rangle 
     = \sqrt{ \frac{2}{3} } \ \Big| (NJ/\psi)_{S_{J/\psi-N}=1/2} N ;  J = 1 \Big\rangle 
           \nonumber  \\
        - \sqrt{ \frac{1}{3} } \ \Big| (NJ/\psi)_{S_{J/\psi-N}=3/2} N ;  J = 1 \Big\rangle. 
   \label{eq:J1T0}
\end{align}
A similar mixing for the $J^{\pi}=1^{-}$, $T=1$ state is given in Appendix A.
Using the decomposition in Eq. (\ref{eq:J1T0}), 
we can determine the contribution of 
spin-independent part, $v_{0}$ and spin-spin part, $v_{s}$ 
of the $J/\psi-N$ potential.
Then in solving this system, 
we can simply take the effective (spin-averaged) potential, $V_{\rm eff}$ 
for $J/\psi-N$ interaction.
For $J^{\pi}=1^{-}$ with $T=0$, $V_{\rm eff}$ is obtained as
\begin{alignat}{6}
 V_{\rm eff}^{(J=1,T=0)}e^{-\mu r^{2}} & \equiv
  \Big\langle (NN)_{S_{NN}=1}, J/\psi ; J = 1  \Big{|} 
  \ v_{J/\psi-N} \ \Big{|} (NN)_{S_{NN}=1}, J/\psi ; J = 1 \Big\rangle \nonumber  \\
    & = ( v_{0} - \frac{1}{2}v_{s} )e^{-\mu r^{2} }. 
\end{alignat}
where $v_{J/\psi-N}$ is given by Eq.(\ref{eq:Jpsi_N_potential2}).
This is regarded as an effective potential between $J/\psi -N$ 
for the $J = 1$ and $T=0$ state.
Note that this expectation value is taken only by the spin part of the wave function 
and the integration of $r$ has not been performed yet.

The similar calculation can be done for the other three channels, 
and after all, in all the four channels with the definite values of $J$, $S_{NN}$ and $T$,
the system is described by a single $J/\psi-N$ effective potential given by 
\begin{equation}
  V_{J/\psi-N}= V_{\rm eff}^{(J, T)}e^{-\mu r^{2} } \label{eq:effective_potential}
\end{equation}
\begin{alignat}{4}
 V_{\rm eff}^{(0,0)} &= v_{0}- v_{s} = v_{\rm eff}(1/2)
\qquad &(J = 0, \ S_{NN}=1,\ T=0) \label{eq:V0Vs1} \\
 V_{\rm eff}^{(1,0)} &= v_{0}- \frac{1}{2}v_{s}     
\qquad &(J = 1,\ S_{NN}=1,\ T=0) \label{eq:V0Vs2} \\
 V_{\rm eff}^{(2,0)} &= v_{0}+ \frac{1}{2}v_{s} = v_{\rm eff}(3/2)
\qquad &(J = 2,\ S_{NN}=1,\ T=0) \label{eq:V0Vs3} \\
 V_{\rm eff}^{(1,1)} &= v_{0}              \qquad &(J = 1,\ S_{NN}=0,\ T=1), \label{eq:V0Vs4}  
\end{alignat}
respectively\footnote{For the details of derivation, see appendix A.}.
Among the $T=0$ channels, the differences of the spin structure 
appear only in the coefficient of the $v_{s}$. 
The binding energy $B$ of the $J/\psi-$deuteron system ($T=0$)
is determined only by the value of $V_{\rm eff}$.
Therefore in the calculations we specify the value of $V_{\rm eff}$, 
and do not change $v_{0}$ and $v_{s}$ separately,
and see how the binding energy of the $J/\psi -$deuteron changes.
%

%
\subsection{$J/\psi-^{4}$He system}
In this section, 
we consider the $J/\psi-^{4}$He bound state. 
This system is suitable for studying spin-independent central part $v_{0}$ 
since the ground state of $^{4}$He is spin 0 and $J/\psi-^{4}$He interaction 
has no contribution from $v_{s}$.
As the binding energy of $^{4}$He is large, 
its wave function may not be disturbed by the relatively weak $J/\psi-N$ interaction.
Therefore we assume that 
the $J/\psi-^{4}$He system can be treated as a two-body system.
As an effective $J/\psi-^{4}$He potential, we use the folding potential 
\begin{alignat}{1}
  V_{\rm fold}(r) = \int v_{J/\psi -N}(\mathbf{r}-\mathbf{r'})\rho(\mathbf{r'}) d^{3}\mathbf{r'}
  \label{eq:Vfold}
\end{alignat}
where $\rho(r)$ is the nucleon density distribution in $^{4}$He. 
We choose a Gaussian function 
\begin{alignat}{1}
  \rho(r) = \rho(0) e^{-(r/b)^{2}}
  \label{eq:rho}
\end{alignat}
with $b = 1.358$ fm, which reproduces the experimental data of 
charge form factor in the elastic electron$-^{4}$He scattering ~\cite{Hofstadter1957,Frosch1967,McCarthy1977}.
Note that the density distribution Eq.(\ref{eq:rho}) is measured from $O$, the center of nucleon 
distribution of $^{4}$He, not from $\bf {R_{G}}$, the center of mass of $^{4}$He. 
But in order to take into account the motion of the center of mass of $^{4}$He, 
we must calculate folding potential from $\mathbf{R_{G}}$. 
We can convert $\rho(r)$ of Eq.(\ref{eq:rho}) measured from $O$ into the distribution 
measured from $\bf R_{G}$ and perform the integral in Eq. (\ref{eq:Vfold}) analytically with 
the $J/\psi-N$ effective potential Eq. (\ref{eq:Jpsi_N_potential2_1}). 
Then we obtain
\begin{alignat}{1}
  V_{\rm fold}(r) = 4\biggl(\frac{4}{4+3\mu b^{2}}\biggr)^{3/2}
                        v_{0} e^{-4\mu r^{2}/(4+3\mu b^{2})}
  \label{eq:Vfold2}
\end{alignat}
for the effective $J/\psi-^{4}$He potential folded from the center of mass of $^{4}$He.
Here $r$ is the relative distance between $J/\psi$ and $^{4}$He measured from $\bf R_{G}$.

\section{Results}
\subsection{Results of $(c\bar{c})-N$ two-body system}
\begin{figure}[!h]
  \begin{minipage}[z]{0.48\linewidth}
  \includegraphics[width=5.0cm, angle=-90, clip, trim=-14 0 0 0]{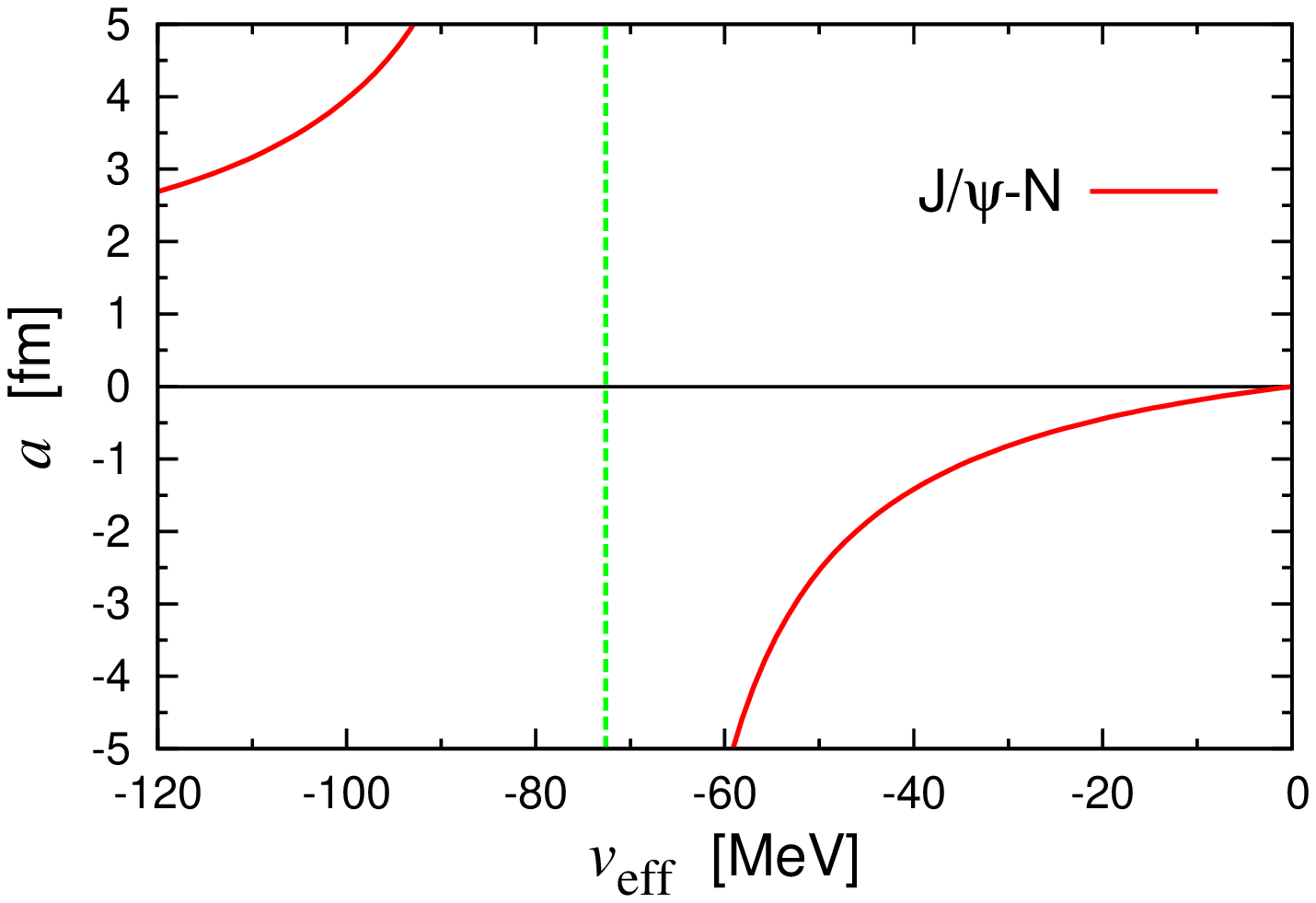}
  \caption{The correspondence of the scattering length $a$ fm 
              and $v_{\rm eff}$ MeV for $J/\psi-N$.}
  \label{fig:JpsiN_10_va}
  \end{minipage}
  \hspace{0.02\linewidth}
  \begin{minipage}[y]{0.48\linewidth}
  \vspace*{-5pt}
  \includegraphics[width= 5.0cm, angle=-90, clip, trim=0 0 -8 0]{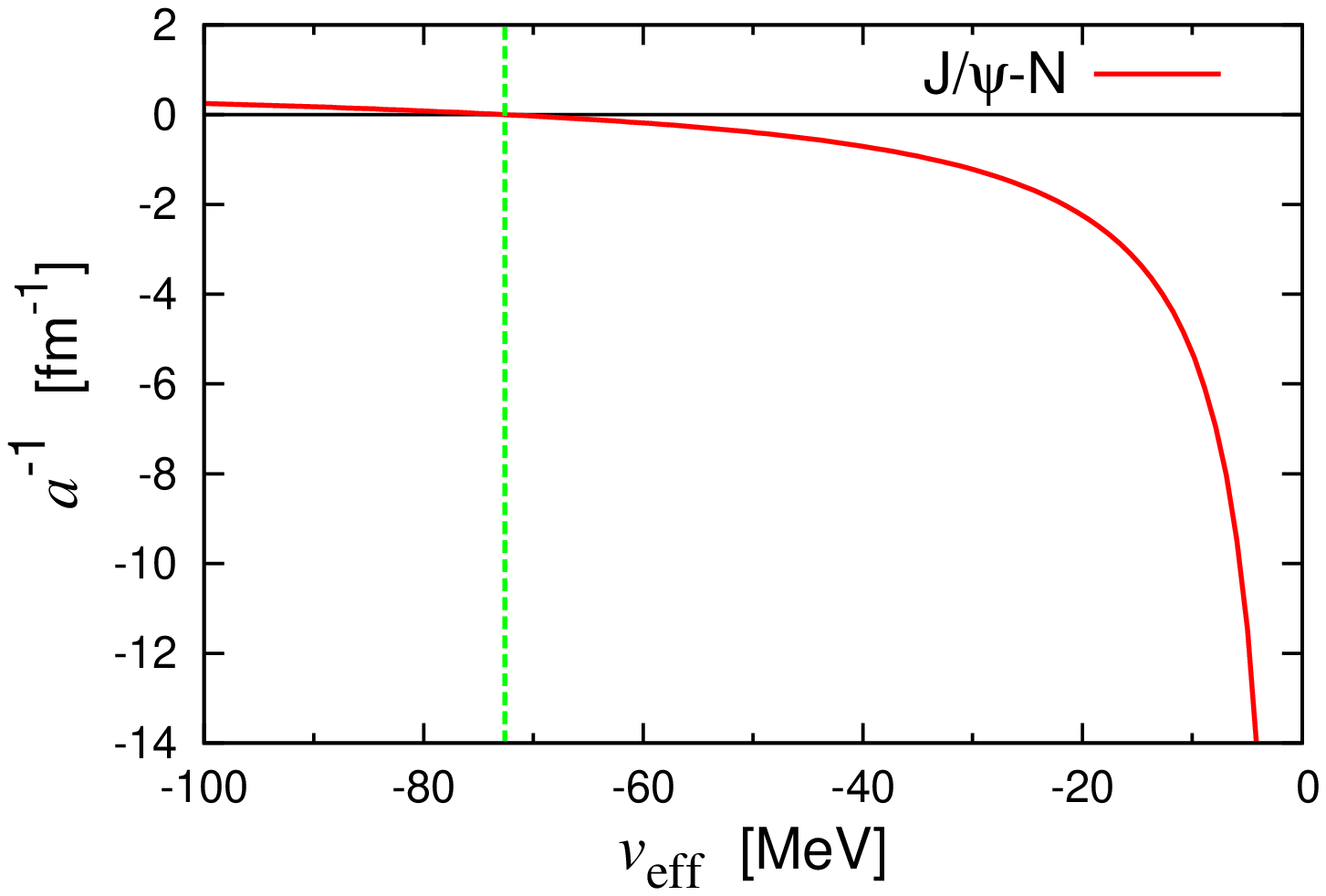}
  \vspace*{-7pt}
  \caption{The correspondence of $a^{-1}$ fm$^{-1}$ 
              and $v_{\rm eff}$ MeV for $J/\psi-N$.}
  \label{fig:JpsiN_10_va_inv}
  \end{minipage}
\end{figure}

The scattering length $a$ of the $J/\psi-N$ system
as a function of $v_{\rm eff}$ in Eq. (\ref{eq:Jpsi_N_potential2}) is shown in Figs. 
\ref{fig:JpsiN_10_va} and \ref{fig:JpsiN_10_va_inv}.
We fix the range parameter of the potential, $\mu=(1.0$ fm)$^{-2}$,
taken from the confinement scale of gluon.
Note that the difference between 
$S_{J/\psi-N}=1/2$ and $3/2$ 
comes merely from the difference of the values of $v_{\rm eff}(S_{J/\psi-N})$. 
Thus the relations between the scattering lengths $a(S_{J/\psi-N})$ 
and $v_{\rm eff}(S_{J/\psi-N})$ for each spin state 
reduce to the same graph as shown in Figs. \ref{fig:JpsiN_10_va} and \ref{fig:JpsiN_10_va_inv}.
We can see that 
the sign of the scattering length $a$ changes at $v_{\rm eff} = -72.6$ MeV.
This value corresponds to the critical strength of the potential 
beyond which there exists a $J/\psi-N$ bound state.
 
A similar relation is given between the scattering length $a$ and $v_{0}$ 
in Eq. (\ref{eq:Etac_N_potential}) 
for the $\eta_{c}-N$ system, 
with $\mu=(1.0$ fm)$^{-2}$.
We find that the sign of the scattering length $a$ changes at $v_{0} = -73.3$ MeV. 
The difference between $\eta_{c}-N$ and $J/\psi-N$ systems 
comes only from the difference of their reduced masses.
%
%
\subsection{$Charmonium-deuteron$ three-body bound states}
\begin{figure}[!h]
  \begin{minipage}[t]{0.46\linewidth}
  \includegraphics[width= 4.8cm, angle = -90, trim = 0 0 0 0]{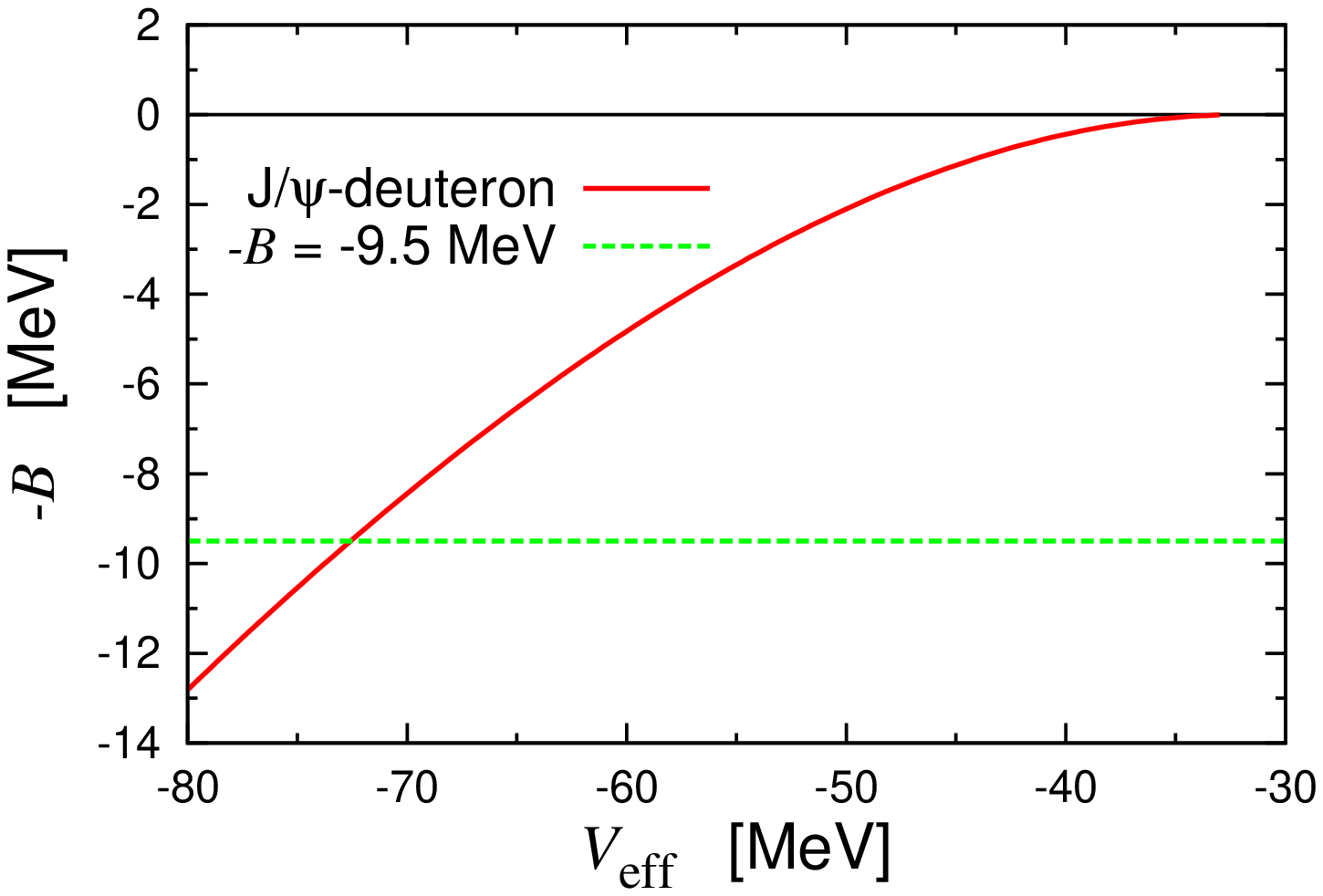}
  \caption{The relation between the binding energy $B$ (MeV) of $J/\psi-$deuteron 
               and $V_{\rm eff}$ (MeV) of $J/\psi -N$ potential.}
  \label{fig:JNN_E_Veff}
  \end{minipage}
   \hspace{0.05\linewidth}
  \begin{minipage}[t]{0.46\linewidth}
  \includegraphics[width= 4.8cm, angle = -90, trim = 0 0 0 0]{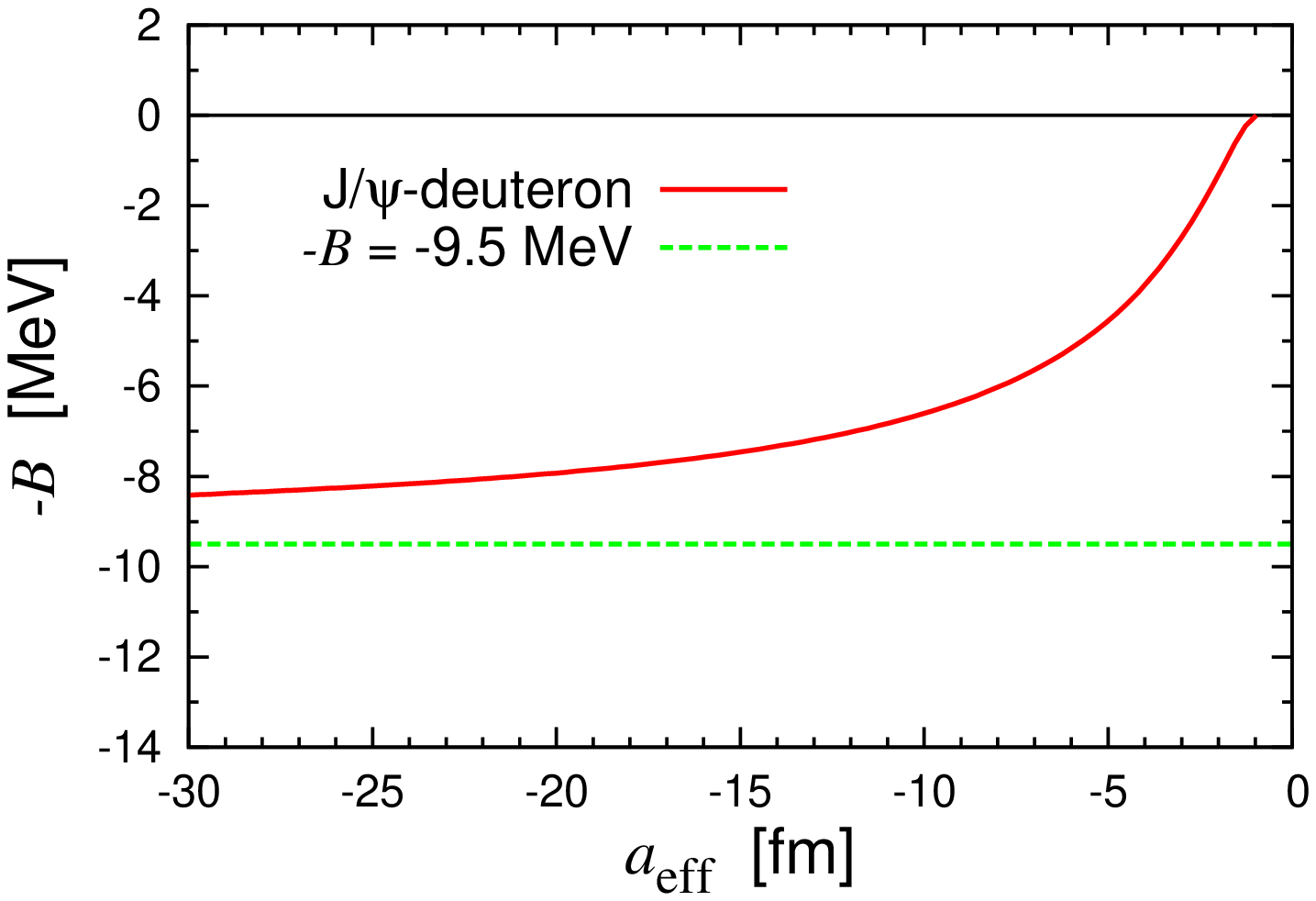}
  \caption{The relation between the binding energy $B$ (MeV) of $J/\psi-$deuteron 
               and the scattering length $a_{\rm eff}$ (fm) of $J/\psi -N$.}
  \label{fig:JNN_aE}
  \end{minipage}
\end{figure}

The relation between $J/\psi-$deuteron binding energy $B$ and 
the potential depth $V_{\rm eff}$ of the effective potential in Eq. (\ref{eq:effective_potential}) ($T=0$)
is shown in Fig. \ref{fig:JNN_E_Veff}.
We fix the range parameter $\mu= (1.0 $ fm$)^{-2}$ as before and 
the binding energy $B$ is measured from deuteron$+ J/\psi$ breakup threshold 
($M_{p} + M_{n} + M_{J/\psi} -2.2$ MeV).
We find that there exists a $J/\psi-$deuteron bound state for $V_{\rm eff}\le -33$ MeV.

We now convert this result into a relation between 
the $J/\psi-N$ scattering length and the $J/\psi-$deuteron binding energy. 
In sect. 3.2, we have shown in Eqs. (\ref{eq:V0Vs1})-(\ref{eq:V0Vs4}) that 
the effective $J/\psi-N$ potential $V_{\rm eff}$ in Eq.(\ref{eq:effective_potential}) 
for a $J/\psi -NN$ system 
is given by a single combination of the spin independent and dependent terms, $v_0$ and $v_s$, 
of the $J/\psi-N$ potential, Eq. (\ref{eq:Jpsi_N_potential2_1}). 
Namely, only one combination appears in the computation of the $J/\psi-NN$ system once the total angular momentum $J$ and isospin $T$ are determined.
Then we define the effective scattering length $a_{\rm eff}$ for each channel, 
corresponding to the $J/\psi-N$ potential given by 
Eqs. (\ref{eq:V0Vs1})-(\ref{eq:V0Vs3}),
and obtain a relation between the $a_{\rm eff}$ and the binding energy, $B$, 
which is shown in Fig. \ref{fig:JNN_aE}. 
In the two-body $J/\psi -N$ calculation (Fig. \ref{fig:JpsiN_10_va}), 
one sees that the scattering length becomes $-\infty$ at $v_{\rm eff} = -72.6$ MeV. 
We find in Fig. \ref{fig:JNN_E_Veff} that the corresponding binding energy of 
the $J/\psi-$deuteron system is $B=9.5$ MeV.
Thus, in Fig. \ref{fig:JNN_aE}, 
the relation approaches $B=9.5$ MeV when $a_{\rm eff}$ approaches $-\infty$.

The effective scattering length $a_{\rm eff}$ 
may not correspond directly to that of the physical $J/\psi -N$ scattering. 
In fact, for $J=0$, we find $V_{\rm eff}^{(J=0,T=0)} = v_0-v_s=v_{\rm eff} (1/2)$, 
and then $a_{\rm eff}$ is 
reduced to $a_{J=1/2}$ for $J/\psi-N$ ($J=1/2$) state. 
Similarly, for $J=2$, we can associate $a_{\rm eff}$ to $a_{J=3/2}$.
In contrast, for $J=1$, there is no one-to-one correspondence 
between $a_{\rm eff}$ and the physical scattering length. 
This is just the scattering length given by the potential Eq. (\ref{eq:V0Vs2}) or (\ref{eq:V0Vs4}), 
which does not correspond to a definite spin of the $J/\psi-N$ system.

\begin{figure}[!h]
  \centering
  \includegraphics[width=5.0cm, angle = -90]{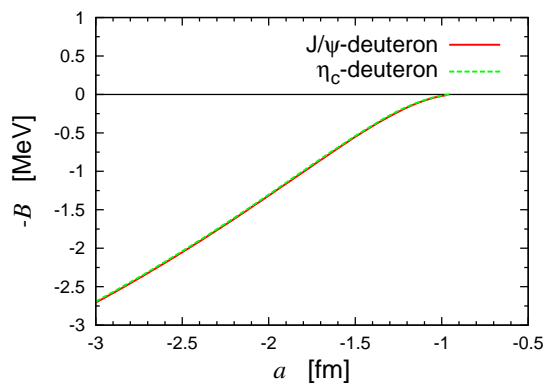}
  \caption{The relation between the binding energy $B$ (MeV) and 
              the scattering length $a$ (fm) for
               $J/\psi-$deuteron and $\eta_{c}-$deuteron.}
  \label{fig:JNN_ENN10_aE}
\end{figure}

Fig. \ref{fig:JNN_ENN10_aE} shows a close-up of the offset of the binding energies of 
the $J/\psi-$deuteron and $\eta_c-$deuteron systems. 
The relation for the $\eta_{c} -$deuteron system $(T=0)$ 
is almost identical to that of $J/\psi -$deuteron $(T=0)$, 
except for the large decay width of $\eta_{c}$ which is not taken into account here 
but will be discussed later.
We find that the critical value of the scattering length 
to have a $J/\psi-$deuteron bound state 
is $-0.95$ fm. 
This is a much stronger attraction than the recent lattice QCD results 
$a \simeq -0.35$ fm \cite{Kawanai2010}, 
which is equivalent to $v_{\rm eff} \simeq -16.7$ MeV.
So there is little possibility of making a $J/\psi-$deuteron bound state 
according to the recent lattice QCD data.
However, it is interesting to see that 
the critical value of the depth of the effective potential 
to have a $J/\psi-$nucleus bound state is reduced from 
$-72.6$ MeV ($A=1$) to $-33$ MeV ($A=2$). 
Thus we expect that the situation may improve for $A \ge 3$ or $4$ and 
there may exist a $J/\psi-$nucleus bound state. 

%
%
\subsection{Conditions on the spin dependent terms}
\begin{figure}[!h]
  \centering
  \includegraphics[width=5cm, angle=-90]{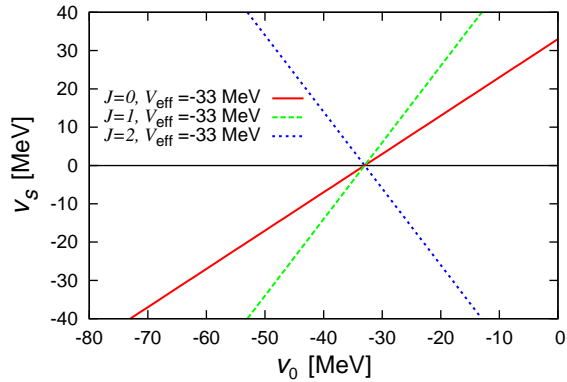}
  \caption{The relations between $v_{0}$, $v_{s}$ and $V_{\rm eff}^{(J,T)}$ for $T=0$.
               The solid (dashed, dotted) line corresponds to the critical line for $J=0 \ (1,2)$, 
               the left side of which is the parameter region for a bound state.}
  \label{fig:V0_Vs_Veff1}
\end{figure}

Finally, we convert the above results on the values of 
the spin-independent ($v_{0}$) and spin-dependent ($v_{s}$) $J/\psi-N$ potentials.
Fig. \ref{fig:V0_Vs_Veff1} summarizes the results. 
The lines going through $(v_{0}, v_{s}) = (-33, 0)$ [MeV] 
are the critical lines for a bound state. 
The left side of the critical line is the region where a bound state exists for each value of $J$. 
In fact, if $v_{s}=0$, i.e., there is no spin dependence in the potential, 
then bound states appear in all $J=0, 1$ and $2$ for $v_{0}$ below $-33$ MeV. 
As a finite $v_{s}$ gives different $V_{\rm eff}$ for $J=0, 1$ and $2$, 
the critical lines differs, for instance, $v_{0} - v_{s} = -33$ MeV for $J=0$ and so on.

We can use these relations to specify the values of $v_{0}$ and $v_{s}$ 
within the model of single Gaussian potential with the same range parameters of 
spin-independent and spin-spin interactions ($\mu = (1.0$ fm$)^{-2}$),
once the binding energies of two spin states, 
for instance, $J=0$ and $1$, are obtained by experiment. 
Each value of the binding energy gives one line on the $v_{0}-v_{s}$ plane 
and two lines with different $J$ will cross at a point. 
This point determines $v_{0}$ and $v_{s}$.

%
\subsection{The range dependence of the binding energy}
\begin{figure}[!h]
  \centering
  \includegraphics[width=5.0cm, angle=-90]{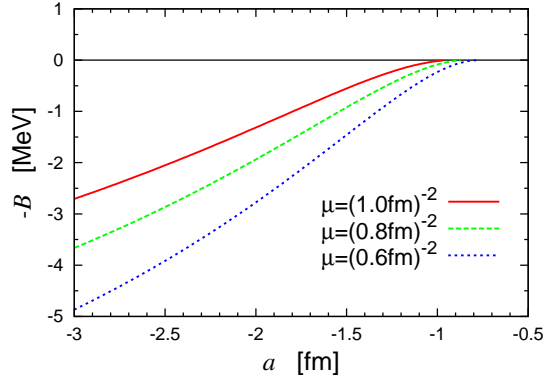}
  \caption{The relations between $(c\bar{c})-$deuteron binding energy $B$ (MeV) and 
              scattering length $a$ (fm) of $(c\bar{c})-N$  
              for $\mu=(1.0$ fm$)^{-2}, (0.8$ fm$)^{-2}$ and $(0.6$ fm$)^{-2}$. }
  \label{fig:Range_dependence1}
\end{figure}

So far we have assumed that 
there is one-to-one correspondence between 
the binding energies of $\eta _{c}-$deuteron and $J/\psi -$deuteron 
and the scattering lengths of $\eta _{c}-N$ and $J/\psi -N$, respectively, 
because the interactions between $\eta _{c}-N$ and $J/\psi -N$
are considered to be short-ranged and weak.
But actually they also depend on the range of the interactions.
Fig. \ref{fig:Range_dependence1} shows the calculations of binding energies 
for the potentials having different values of range parameters, i.e., 
$\mu=(1.0$ fm$)^{-2}, (0.8$ fm$)^{-2}$ and $(0.6$ fm$)^{-2}$.
The results are almost the same for 
both the $\eta _{c}-$deuteron and $J/\psi -$deuteron cases.
It shows that a smaller potential range gives a deeper binding energy 
when the values of the scattering lengths are the same. 
The difference of the binding energy grows larger 
as the absolute value of the scattering length (i.e. the corresponding attraction) increases. 
The difference becomes small when $|a|$ and corresponding attraction decreases.
In case of $\mu=(0.6$ fm$)^{-2}$, $a \le -0.79$ fm is needed to make a bound state,
which is slightly above from $a \le -0.95$ fm in case of $\mu=(1.0$ fm$)^{-2}$.
\begin{figure}[!h]
  \begin{minipage}[t]{0.47\linewidth}
  \includegraphics[width=5.0cm, angle=-90]{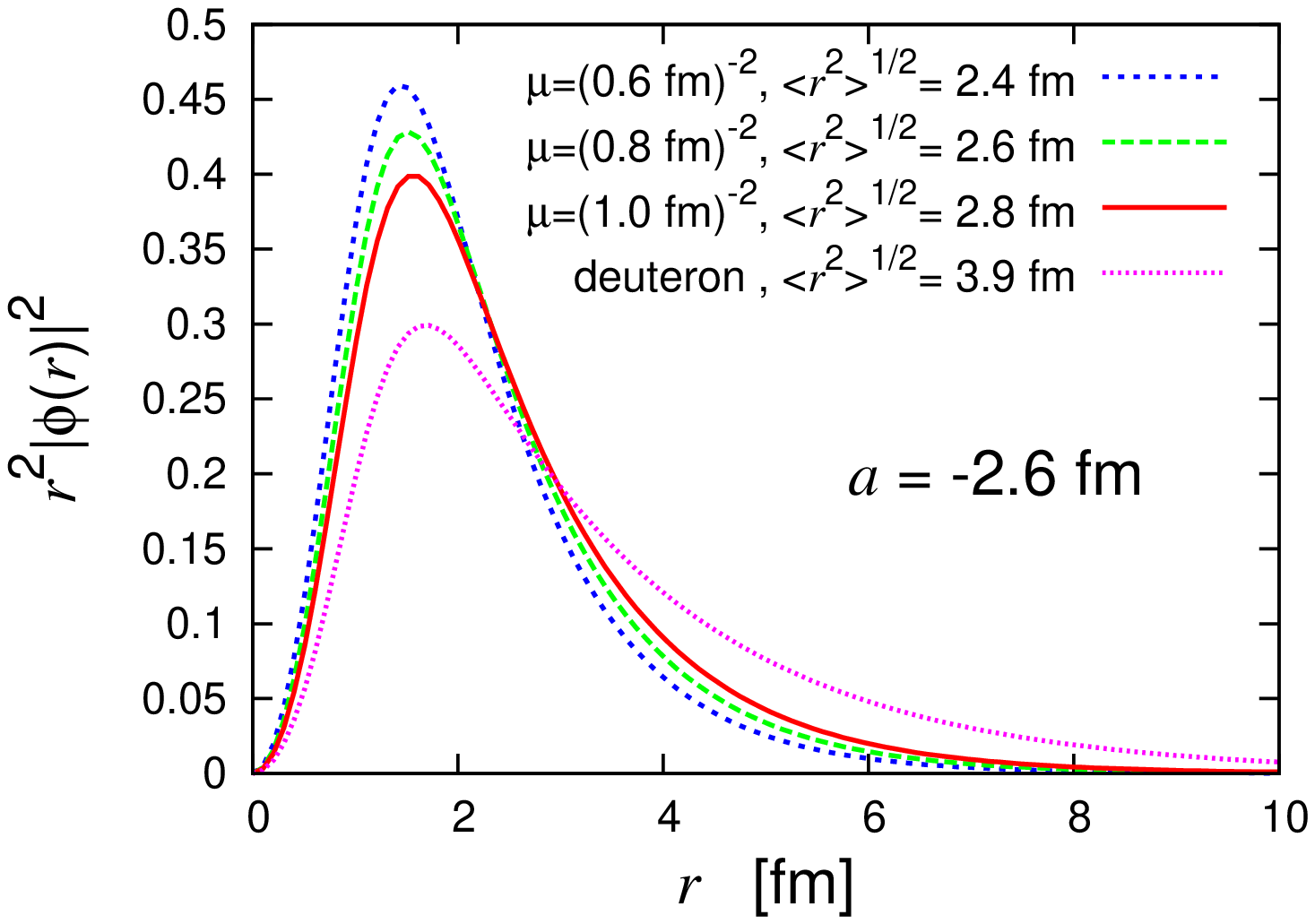}
  \caption{Density distributions 
              between $p-n$ in $(c\bar{c})-NN$ system 
              for $\mu=(1.0$ fm$)^{-2}, (0.8$ fm$)^{-2}$ and $(0.6$ fm$)^{-2}$, 
              each corresponding to $a=-2.6$ fm. 
              The density distribution of the deuteron is also shown.}
  \label{fig:Range_dependence2}

  \end{minipage}
  \hspace{0.05\linewidth}
  \begin{minipage}[t]{0.47\linewidth}
  \includegraphics[width=5.0cm, angle=-90]{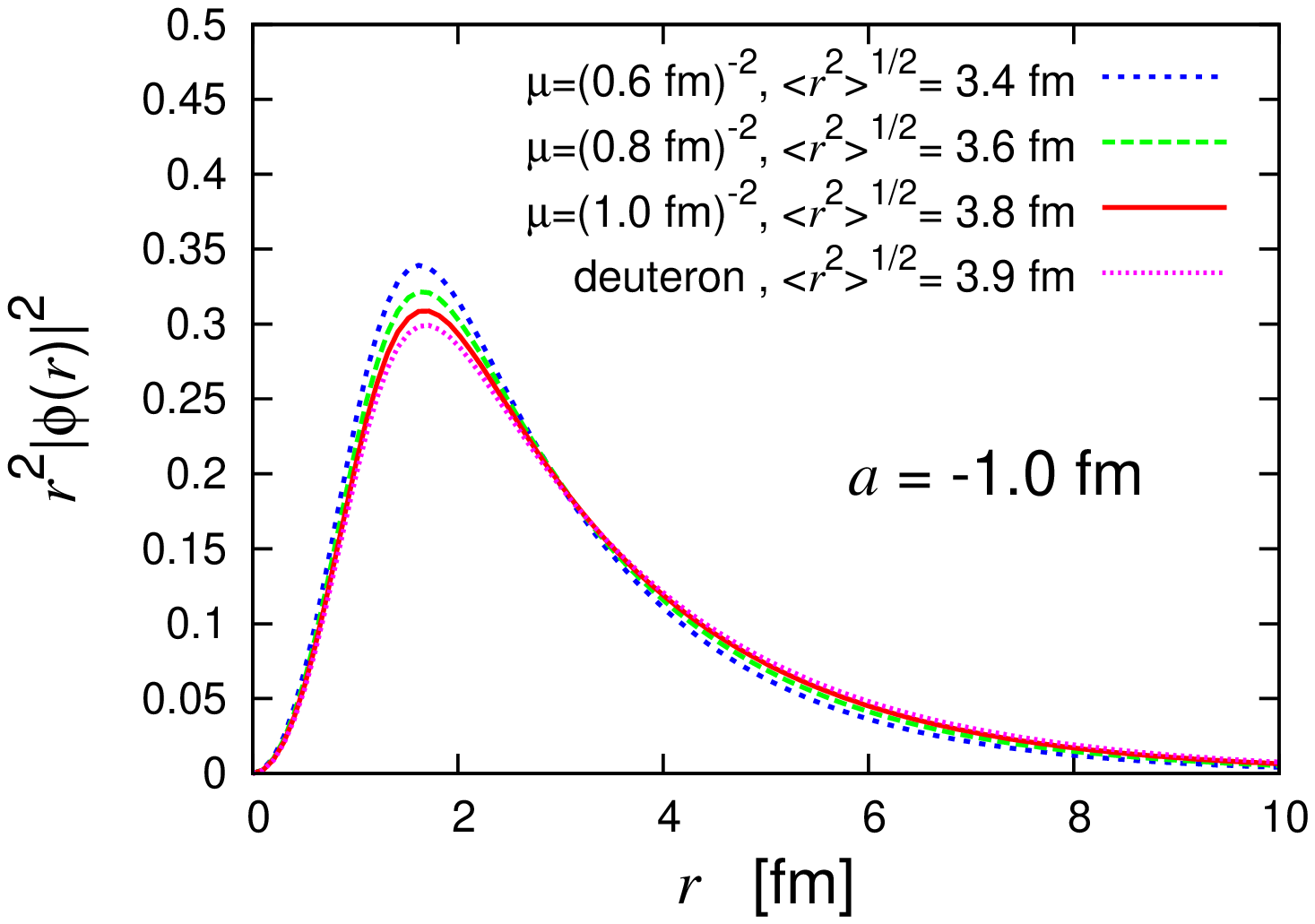}
  \caption{Density distributions 
              between $p-n$ in $(c\bar{c})-NN$ system 
              for $\mu=(1.0$ fm$)^{-2}, (0.8$ fm$)^{-2}$ and $(0.6$ fm$)^{-2}$, 
              each corresponding to $a=-1.0$ fm. 
              The density distribution of the deuteron is also shown.}
  \label{fig:Range_dependence3}
  \end{minipage}
\end{figure}

These features can well be understood if we look at the density distribution between $p-n$ 
in the $(c\bar{c})-NN$ system. 
In Fig. \ref{fig:Range_dependence2}, 
density distributions between $p-n$ in the $(c\bar{c})-NN$ system 
in the cases of $\mu=(1.0$ fm$)^{-2}, (0.8$ fm$)^{-2}$ and $(0.6$ fm$)^{-2}$,  
are shown together with that of the deuteron. 
$r$ denotes the relative distance between $p$ and $n$. 
The strength of the $(c\bar{c})-N$ potential is fixed so that it corresponds to $a = -2.6$ fm, 
the same value as used in \cite{Wasson1991}. 
The deuteron wave function is calculated by 
the Minnesota potential
\cite{Reichstein1970, Thompson1977}. 
The corresponding binding energies $B$ of the $(c\bar{c})-NN$ system 
measured from the $(c\bar{c})+$deuteron breakup threshold are 
$2.2$ MeV, $3.0$ MeV and $4.2$ MeV and 
the root mean square distances are $2.8$ fm, $2.6$ fm and $2.4$ fm for 
$\mu=(1.0$ fm$)^{-2}, (0.8$ fm$)^{-2}$ and $(0.6$ fm$)^{-2}$ cases, respectively.
We can see in the Fig. \ref{fig:Range_dependence2} that 
the $p-n$ density distribution shrinks by the emergence of $(c\bar{c})$ 
and it shrinks more when the range of the $(c\bar{c})-N$ potential is shorter. 
This is because in order to give the same value of the scattering length 
with shorter potential range, the depth of the potential must be deeper 
and then the effect of $(c\bar{c})$ attracting nucleons grows rapidly 
at short distances. 

%
\subsection{The difference between using Gaussian and Yukawa-type potentials}
\begin{figure}[!h]
  \centering
  \includegraphics[width=5.0cm, angle=-90]{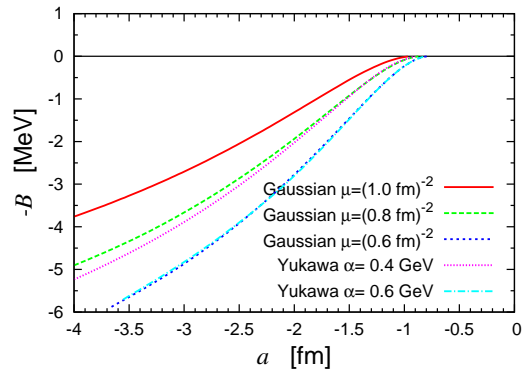}
  \caption{The relations between $(c\bar{c})-$deuteron binding energy $B$ (MeV) and 
              scattering length $a$ (fm) of $(c\bar{c})-N$  
              for Gaussian potentials with $\mu=(1.0$ fm$)^{-2}, (0.8$ fm$)^{-2}$ and $(0.6$ fm$)^{-2}$
             and for Yukawa-type potentials with $\alpha = 0.4$ and $0.6$ GeV. }
  \label{fig:potential_form_dependence}
\end{figure}

Another question is that whether our results would be changed by using the other potential forms, 
for example, Yukawa-type potential (Eq. (\ref{eq:YukawaPotential})). 
In Fig. \ref{fig:potential_form_dependence}, 
we show the relations between $(c\bar{c})$-deuteron binding energy $B$ and 
the $(c\bar{c})-N$ scattering length $a$ 
which are calculated by using Yukawa-type potential 
with the range parameters $\alpha = 0.4$ and $0.6$ GeV in Eq. (\ref{eq:YukawaPotential}), 
together with the results using Gaussian potentials (the same as Fig. \ref{fig:Range_dependence1}). 
We see in the figure that the difference between using Gaussian and Yukawa-type potentials 
is small.
Especially, the curve of $\mu = (0.6$ fm$)^{-2}$ for Gaussian potential 
is almost identical to that of $\alpha = 0.6$ GeV for Yukawa-type potential.
Therefore we conclude that 
our results do not change qualitatively by the choice of the potential form. 

Wasson \cite{Wasson1991} used the Yukawa-type potential Eq.(\ref{eq:YukawaPotential}) 
with the values of parameters corresponding to $a=-2.6$ fm for $\eta_{c}-N$ system 
(and $a=-2.8$ fm for $J/\psi - N$ system). 
They found no bound state in the $(c\bar{c})-NN$ (A=2) system, 
which seems inconsistent with our results (Fig. \ref{fig:potential_form_dependence}).
One of the reasons could be that 
they used a folding potential which is folded from the center of the nucleon distribution, 
not from the center of mass, and did not implement the center of mass correction (CMC), 
i.e., did not remove the center of mass motion. 
We will see the difference between implementing and not implementing CMC later 
in the case of $(c\bar{c})-^{4}$He. 
The other reason could be that they used a folding potential with a fixed nucleon density 
and ignored the effect of nuclear shrinkage.
As can be seen in Figs. \ref{fig:Range_dependence1} and \ref{fig:Range_dependence2}, 
the effect of nuclear shrinkage cannot be ignored 
for such a strong $(c\bar{c})-N$ attraction with $a=-2.6$ fm. 
In light and loosely bound nuclei such as deuteron, 
the shrinkage effect becomes negligible only when the attraction is weaker than 
at least $a \gtrsim -1.0$ fm, 
as can be seen from Figs. \ref{fig:Range_dependence1} and \ref{fig:Range_dependence3}.

In summary, together with the discussions of subsection 4.4,  
we see that there are not only the scattering length dependence but also 
the range dependence of the binding energy. 
But when the attraction is weak, the range dependence is small. 
Also, the choice of the potential form does not affect the results seriously. 
%
%
\subsection{$J/\psi-^{4}$He}
\begin{figure}[!h]
  \begin{minipage}[t]{0.45\linewidth}
  \includegraphics[width=0.68\linewidth, angle=-90]{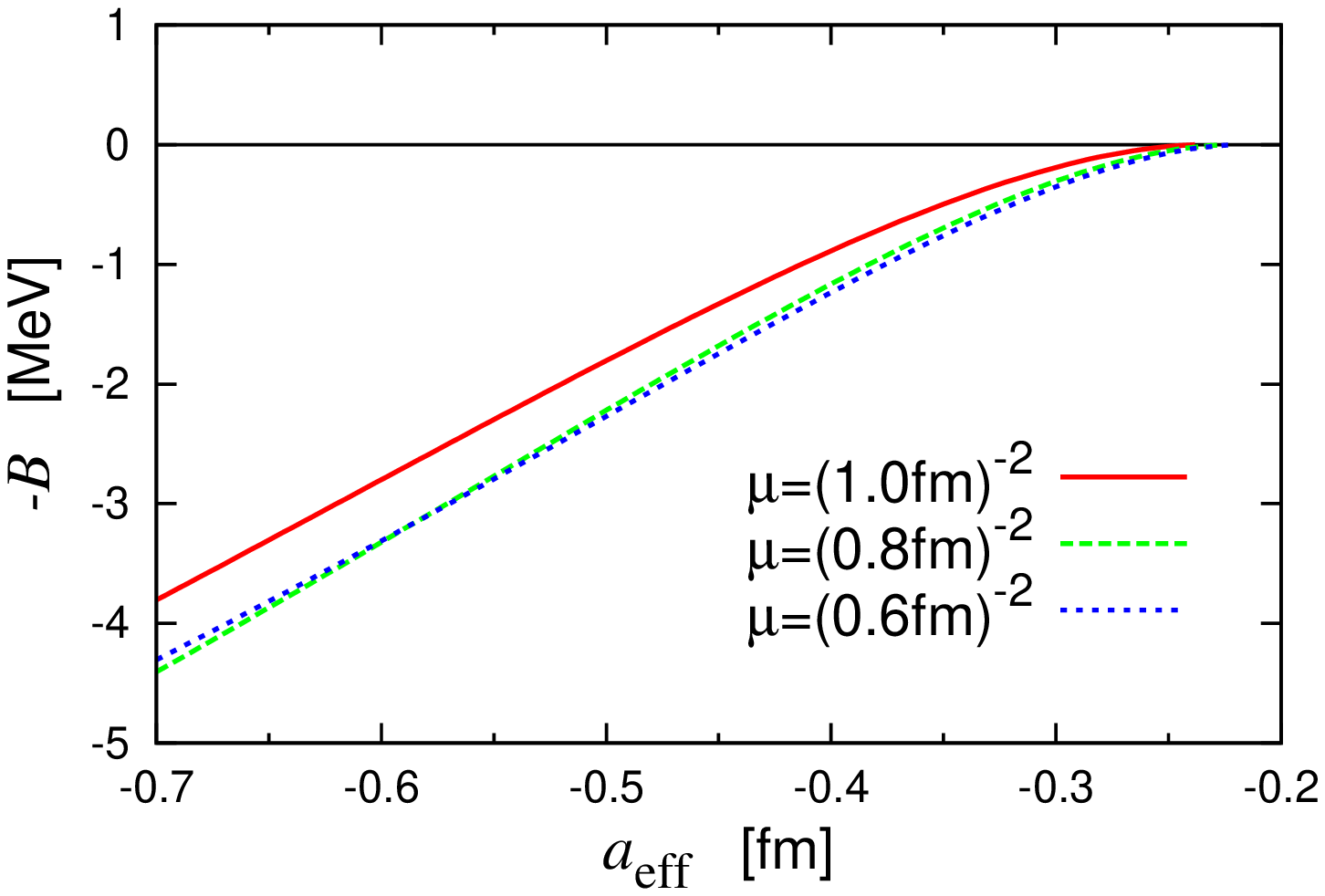}
  \caption{The relation between the binding energy $B$ of $J/\psi -^{4}$He 
              and the $J/\psi -N$ scattering length $a_{\rm eff}$.}
  \label{fig:J_4He_E_a}
  \end{minipage}
  \hspace{0.05\linewidth}
  \begin{minipage}[t]{0.45\linewidth}
  \includegraphics[width=0.7\linewidth, angle=-90]{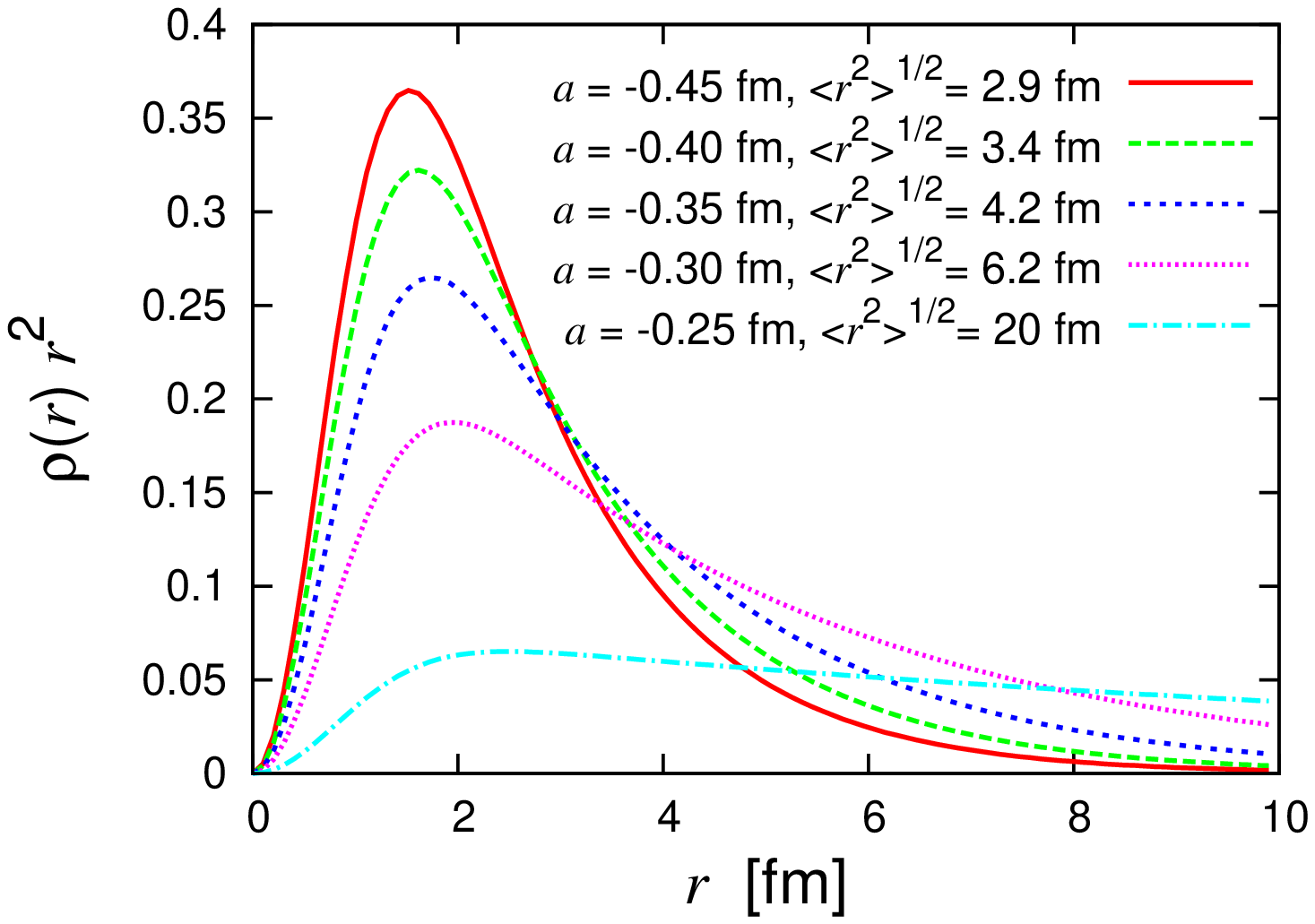}
  \caption{The density distribution $\rho(r)r^{2}$ between $J/\psi$ and $^{4}$He 
              with $\mu=(1.0$ fm)$^{-2}$.}
  \label{fig:J_4He_density}
  \end{minipage}
\end{figure}

The relation between $J/\psi-^{4}$He binding energy $B$ and 
the $J/\psi-N$ scattering length $a_{\rm eff}$ 
is shown in Fig. \ref{fig:J_4He_E_a}. 
The range parameter $\mu$ of $J/\psi-N$ potential 
is chosen to be $\mu = (0.6$ fm)$^{-2}, (0.8$ fm)$^{-2}$ and $(1.0$ fm)$^{-2}$. 
We can see the range dependence of $B$ becomes small when $|a_{\rm eff}|$ becomes small.
A $J/\psi -^{4}$He bound state could be formed when 
$a_{\rm eff} \le -0.24$ fm for all cases. 
The critical values of $a_{\rm eff}$ for $\mu = (0.8$ fm)$^{-2}$ and $(0.6$ fm)$^{-2}$ 
are shifted by about $\sim 0.02$ fm. 
The binding energies are, for example, when $a_{\rm eff}=-0.35$ fm, 
$B = 0.50, 0.69$ and $0.76$ MeV for $\mu = (1.0$ fm)$^{-2}, (0.8$ fm)$^{-2}$ 
and $(0.6$ fm)$^{-2}$, respectively.
Comparing with the lattice QCD data\cite{Kawanai2010}, $a_{J/\psi-N} \simeq -0.35$ fm, 
this result supports the existence of a shallow $J/\psi-^{4}$He bound state. 

Here it should be noted that 
the net spin-spin interactions between the nucleon and $J/\psi$ in $J/\psi-^{4}$He system 
cancel out so that 
the contribution only from the spin-independent part, 
$v_{0}$ of Eq. (\ref{eq:Jpsi_N_potential2_1}), 
comes into account. 
Therefore 
the scattering length $a_{\rm eff}$ corresponds to the spin-independent part, $v_0$.

The density distribution $\rho(r)r^{2}$ between $J/\psi$ and $^{4}$He 
with $\mu=(1.0$ fm)$^{-2}$ 
is shown in Fig. \ref{fig:J_4He_density}, 
where $r$ is the relative coordinate of $J/\psi$ and the center of mass of $^{4}$He. 
The value of $v_{0}$ is varied from $-12.7$ to $-20.1$ MeV, 
which correspond to the scattering lengths $a_{\rm eff}$ from $-0.25$ to $-0.45$ fm.
For example, 
$v_{0}=-14.8,-16.7$ and $18.5$ MeV corresponds to $a_{\rm eff}=-0.30,-0.35$ and $-0.40$ fm.
We see that they have a long tail while having peaks around $r=1.5 \sim 2.0$ fm. 

Comparing our results with that of Wasson \cite{Wasson1991}, there is some discrepancy. 
While their binding energy of $(c\bar{c})-^{4}$He is $B = 5.0$ MeV, 
our result corresponding to the potential for $a=-2.6$ fm used in \cite{Wasson1991} 
is $B = 15.7$ MeV. 
This difference seems due to the effect of the treatment of the center of mass motion.
Wasson used a potential 
folded from the center of nucleon distribution in nucleus, not from the center of mass. 
But in light nuclei, such as $^{4}$He, the motion of the center of mass is not negligible that 
we must correctly fold the potential from the center of mass. 

\begin{figure}[!h]
  \centering
  \includegraphics[width=6cm, angle=-90]{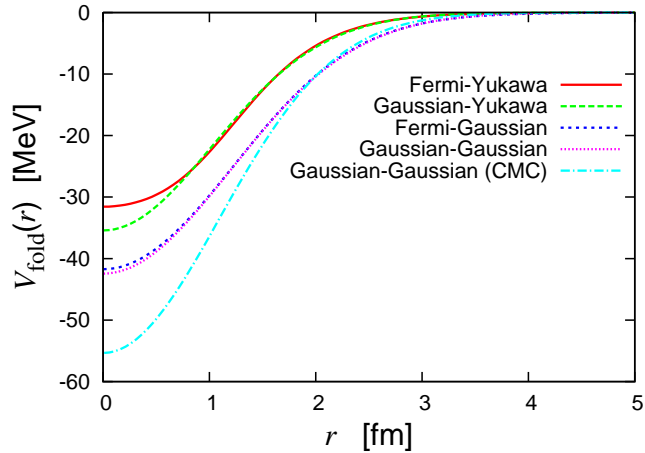}
  \caption{The $(c\bar{c})-^4$He folding potentials 
              with different density distributions and potentials 
              for (1) Fermi-Yukawa, (2) Gaussian-Yukawa, (3) Fermi-Gaussian, 
                  (4) Gaussian-Gaussian and 
                  (5) Gaussian-Gaussian with center of mass correction (CMC).
              The relative distance $r$ is measured 
              from the center of nucleon distribution of $^{4}$He for the first four lines and 
              from the center of mass of $^{4}$He for the last line.}
  \label{fig:folding1}
\end{figure}

In Fig. \ref{fig:folding1}, we show 
several $(c\bar{c})-^4$He folding potentials using different density distributions and potentials 
and folded from the different origin points. 
The two types of nucleon density distributions are 
the Fermi three-parameter distribution, Eq. (9) of \cite{McCarthy1977}, 
with $c=0.964$ fm, $z =0.322$ fm, $\omega=0.517$ and $\rho_{0}=0.05993$ 
and the Gaussian type distribution, Eq. (\ref{eq:rho}), with $b=1.358$ fm. 
The $(c\bar{c})-N$ potentials are 
the Yukawa type, Eq. (\ref{eq:YukawaPotential}), with $A=-0.6$ and $\alpha=600$ MeV and 
the Gaussian type, Eq. (\ref{eq:GaussPotential}), with $v_{\rm eff}=-50.9$ MeV
and $\mu=(1.0$ fm$)^{-2}$, 
both corresponding to the scattering length $a= -2.6$ fm for $\eta_{c}-N$.
The first four lines are folded from the center of nucleon distribution of $^{4}$He, 
not from the center of mass. 
The Wasson's potential corresponds to the first line (see also the Fig. 1 of \cite{Wasson1991}).
The last line is the potential folded from the center of mass of $^{4}$He 
using Gaussian distribution and Gaussian potential. 
The relative distance $r$ between $(c\bar{c})$ and $^{4}$He is measured from 
the center of nucleon distribution of $^{4}$He for the first four lines folded from the point 
and 
for the last line 
from the center of mass of $^{4}$He. 
The figure shows that the effects of folding from different origin points 
are quite large in $^{4}$He, giving different binding energies, $B = 5.0$ MeV and $B = 15.7$ MeV.
The difference between using the Fermi three-parameter distribution and 
the Gaussian distribution is not so large compared to the difference of using different potentials.
Therefore, we conclude that it is important to 
correctly fold the potential from the center of mass of nucleus, 
not from the center of nucleon distribution, 
in case of light nucleus such as $^{4}$He.

\section{$J/\psi - \eta_{c}$ mixing and the decay of charmonium in nuclei}
So far we have neglected the decay and the mixing of $J/\psi$ and $\eta_{c}$ in nuclei. 
Let us discuss briefly about them in this section and show that 
our results obtained above are not seriously affected by them, 
especially for the $J/\psi-$nucleus bound states. 

The $J/\psi$ has the spin $J^{\pi} = 1^{-}$, the mass $m_{J/\psi} = 3096.9 \pm 0.0$ MeV, and 
the full decay width $\Gamma = 92.9 \pm 2.8$ keV,
while 
$\eta_{c}$ has $J^{\pi} = 0^{-}$, $m_{\eta_{c}} = 2981.0 \pm 1.1$ MeV 
and $\Gamma = 29.7 \pm 1.0$ MeV\cite{PDG}.
The large decay width  of $\eta_{c}$ will indicate that $\eta_{c}$ is unstable in nuclei.  
On the other hand, the width of $J/\psi$ is small enough that we expect  
narrow $J/\psi$ states in nuclei. 

Two possible mechanisms, which make the decay width of charmonium($J/\psi$) nuclear states larger are  
(1) the final state interaction of charmonium decay products with nucleon in nucleus, and 
(2) the mixing of charmonium-nucleon state with the other hadronic states which retain $c$ and $\bar{c}$ quarks. 

In the case (1) for $J/\psi$, the final state interaction in nucleus,
for example, absorption of $\pi$ made from $J/\psi$ decay by nucleon in nucleus, 
may enhance the decay width of $J/\psi$ in nucleus several times larger than in vacuum. 
But the decay width of $J/\psi$ in vacuum is so small ($\simeq 93$ keV) that 
even if it is enhanced for several times larger in nucleus, it still would be small ($\le 1$ MeV). 

In the case (2), the decay of $J/\psi$ going through the mixing with the other hadronic states 
which divide $c$ and $\bar{c}$ quarks separately to the hadrons are prohibited 
since such hadronic states have larger masses. 
For example, the lightest charmed meson and baryon are $D$ and $\Lambda_{c}$.
But the mass of the lightest state 
which has the same quark components as $p + J/\psi$ (or $p + \eta_{c}$),  
$m_{\Lambda_{c} } + m_{\bar{D}^{0} } = 4151.3 $ MeV, 
is much heavier than $m_{p} + m_{J/\psi } = 4035.2 $ MeV (and $m_{p} + m_{\eta_{c} } = 3919.3$ MeV). 
Therefore $J/\psi$ (and $\eta_{c}$) in nucleus 
cannot decay by the strong interaction via mixing of these states. 

Then, the only possible decay process of $J/\psi$ in this case goes through the mixing of 
$J/\psi -$nucleus and $\eta_{c}-$nucleus channels 
which have the same conserving quantum numbers. 
The mixing process can further be divided into two groups.  
One is the coherent mixing which retains the nucleus to its ground state in the mixing process, 
for example,  
$J/\psi - $deuteron $(J^{\pi}=1^{-}) \rightarrow \eta_{c} - $deuteron $(J^{\pi}=1^{-}, L = 0)$. 
The other is the incoherent mixing in which the nucleus is excited (or broken),  
e.g., $J/\psi - ^{4}$He $ \rightarrow \eta_{c} - ^{3}$H$- p$.

\subsection{The $J/\psi - \eta_{c}$ coherent mixing in nuclei }
Since $J/\psi$ and $\eta_{c}$ have the same quark components $(c\bar{c})$, 
it is possible that $J/\psi -$nucleus and $\eta_{c} -$nucleus states are mixed with each other
when their conserving quantum numbers coincide. 
The mixing, however, is suppressed in the heavy quark regime, because the mixing requires
the spin-flipping interactions of the charm quark. Such interactions are proportional 
to the inverse of the (heavy) quark mass and are thus will be suppressed.

Furthermore, for the coherent mixing,
$J/\psi-^{4}$He and $J = 0$ channel of $J/\psi-$deuteron 
cannot have the same $J$ with $\eta_{c}-^{4}$He and $\eta_{c}-$deuteron, 
respectively, as can be seen in Table \ref{fig:charmonium_nucleus_spin_states} 
which lists the coherent mixing channels of  
$J/\psi-$nucleus and $\eta_{c}-$nucleus for the nuclei used in the present paper.
\begin{table}[!h]
    \centering
    \begin{tabular}{ccccccc} \hline 
                                &    $J$       &                         &                               & $L$ &              
 \\
     \hline \addlinespace[2pt]
     $J/\psi-N$           &  $1/2^{-}$ &  $\rightarrow$   & $\eta_{c}-N$            &  0    &                   
 \\ 
                                &  $3/2^{-}$ &  $\rightarrow$   & $\eta_{c}-N$            &  2    &                
 \\
     \addlinespace[2pt]  \hline \addlinespace[3pt]
   $J/\psi-$deuteron   &  $0^{-}$    &  $\rightarrow$   &       no                    &        &                
 \\
                                &  $1^{-}$    &  $\rightarrow$  &  $\eta_{c}-$deuteron  & $0, 2$  &        
 \\
                                &  $2^{-}$    &  $\rightarrow$   &  $\eta_{c}-$deuteron  & $2$   &              
 \\
     \addlinespace[3pt] \hline \addlinespace[2pt]  
     $J/\psi-^{4}$He    &  $1^{-}$    &  $\rightarrow$   &        no                  &        &       
 \\ 
     \addlinespace[3pt] \hline
    \end{tabular}
    \caption{Coherent mixing channels of
                $J/\psi -$nucleus and $\eta_{c}-$nucleus systems for several light nuclei.
                $L$ denotes the orbital angular momentum between charmonium and nucleus.}
    \label{fig:charmonium_nucleus_spin_states}
\end{table} 

By all the discussions in this section, 
we conclude that  
the nuclear medium effect of the charmonium ($J/\psi$) decay 
and the mixing of $J/\psi$ and $\eta_{c}$ in light nuclei can be neglected to the leading order in $1/m_Q^2$.

\section{Summary and conclusion}
In this work, we have introduced effective Gaussian potentials 
for the $(c\bar{c})-N$ interaction and obtained the relations between the scattering lengths $a$
and the strength parameters, $v_{0}$ (Eq.(\ref{eq:Etac_N_potential})) for $\eta_{c}-N$  
and $v_{\rm eff}$ (Eq.(\ref{eq:Jpsi_N_potential2})) for $J/\psi-N$, 
of the effective potential.
Then we have examined possibilities of bound $(c\bar{c})-$deuteron ($NN$) 
and $(c\bar{c})-^{4}$He systems. 
The relations between binding energies, $B$, 
and potential strength, $v_{0}$ for $\eta_{c}-$deuteron 
and $V_{\rm eff}$ (Eq.(\ref{eq:effective_potential})) for $J/\psi -$deuteron, 
are given by solving the Schr\"odinger equation with GEM.
Combining these results we obtain the relations between $B$ and the scattering length $a$ 
for both $\eta_{c}-$deuteron and $J/\psi -$deuteron cases. 
Both for $\eta_{c}-$deuteron and $J/\psi -$deuteron, 
$a \le -0.79$ fm ($\mu=(0.6$ fm$)^{-2}$) or 
$a \le -0.95$ fm ($\mu=(1.0$ fm$)^{-2}$) is needed to make a bound state.
Comparing with the lattice QCD data 
$a_{\eta_{c}-N} \sim -0.25$ fm and $a_{J/\psi-N} \sim -0.35$ fm~\cite{Kawanai2010}, 
the obtained results indicate that 
it is unlikely for the $A \le 2$ nuclei to make a bound state with charmonia
 $\eta_{c}$ and $J/\psi$.
We have also checked that the range dependence of the binding energy decreases 
as the attractions become weaker.

For the calculation of the binding energy of $J/\psi-^{4}$He system, 
we employ the folding potential.
The result shows that $a \le -0.24$ fm is needed to make $J/\psi-^{4}$He bound state.
Thus, the value of the scattering length obtained from lattice QCD data 
$a_{J/\psi-N} \sim -0.35$ fm~\cite{Kawanai2010}
is large enough to make a $J/\psi-^{4}$He bound state. 
The binding energy is $B \simeq 0.5$ MeV for $a_{J/\psi-N} \sim -0.35$ fm.
We here note that as far as the binding energy is small, 
the nucleonic wave function of $^{4}$He may not be deformed by the $J/\psi$ binding. 
Then the use of the folding potential is justified.

In conclusion, we find from a simple effective potential analyses that 
the charmonium ($c\bar{c}$) may form bound states 
in the nuclei of $A \ge 4$, supposing that 
the current lattice QCD evaluation of the charmonium-nucleon scattering lengths are reliable.

%

\section*{Acknowledgment}
The authors would like to thank 
Prof. Ryoichi Seki, Shoichi Sasaki and Dr. Taichi Kawanai for useful discussions and information.
This work is supported 
in part by the Grant-in-Aid for Scientific Research 
from MEXT and JSPS (Nos. 24540294, 22105503).
A. Y. acknowledges the financial support from the Global 
Center of Excellence Program by MEXT, Japan through the 
"Nanoscience and Quantum Physics" Project of the Tokyo 
Institute of Technology.

\appendix
\section{Spin matrix elements of $J/\psi-NN$ potential}
In this appendix, we derive the relevant combinations of the $J/\psi-N$ potential 
for the $A=2$ $J/\psi-NN$ system.
\begin{enumerate}
\item $J=1$, $S_{NN}=1$  ($T=0$)
\begin{align}
   \Big| (NN)_{S_{NN}=1} J/\psi ; J=1 \Big\rangle 
    &= \sqrt{ \frac{2}{3} } \ \Big| (NJ/\psi)_{S_{J/\psi-N}=1/2} N ;  J = 1 \Big\rangle  
      \nonumber \\ 
    &- \sqrt{ \frac{1}{3} } \ \Big| (NJ/\psi)_{S_{J/\psi-N}=3/2} N ;  J = 1 \Big\rangle 
\end{align}
We define the expectation value of the potential 
taken by the spin-wave function as $V_{\rm eff}^{(J,T)}e^{-\mu r^{2}}$.
Since two states in different values of $S_{J/\psi -N}$ are orthogonal,
we obtain
\begin{alignat}{6}
  \Big\langle &(NN)_{S_{NN}=1} J/\psi ; J = 1 \ \Big| \ 
                   v_{J/\psi-N} \ \Big| (NN)_{S_{NN}=1} J/\psi ; J = 1 \Big\rangle \nonumber \\
     &=  \frac{2}{3} \ \Big\langle (NJ/\psi)_{S_{J/\psi-N}=1/2} N ;  J = 1 \ \Big| \ 
          v_{J/\psi-N} \ \Big| (NJ/\psi)_{S_{J/\psi-N}=1/2} N ;  J = 1 \Big\rangle  
              \nonumber  \\ 
 & \qquad \quad+ \frac{1}{3} \ \Big\langle (NJ/\psi)_{S_{J/\psi-N}=3/2} N ;  J = 1 \ \Big| \ 
          v_{J/\psi-N} \ \Big| (NJ/\psi)_{S_{J/\psi-N}=3/2} N ;  J = 1 \Big\rangle   
              \nonumber  \\ 
     &= \frac{2}{3} \ \Big[ (v_{0} - v_{s})e^{-\mu r^{2}} \Big]
       + \frac{1}{3} \ \Big[ (v_{0} + \frac{1}{2} v_{s})e^{-\mu r^{2}} \Big]  \nonumber  \\ 
     &= \ ( v_{0} - \frac{1}{2}  v_{s} )e^{-\mu r^{2} }. 
\end{alignat}
Thus, we obtain 
\begin{alignat}{1}
   V_{\rm eff}^{(J=1,T=0)} = \ v_{0} - \frac{1}{2} v_{s} \ .
\end{alignat}

\item $J=1$, $S_{NN}=0$ ($T=1$)

Using the spin decomposition, 
\begin{align}
   \Big| (NN)_{S=0} J/\psi ; J = 1 \Big\rangle 
  &= \sqrt{ \frac{1}{3} } \ \Big| (NJ/\psi)_{S=1/2} N ;  J = 1 \Big\rangle   \nonumber \\ 
  &+ \sqrt{ \frac{2}{3} } \ \Big| (NJ/\psi)_{S=3/2} N ;  J = 1 \Big\rangle
\end{align}
we obtain 
%
\begin{alignat}{5}
  \Big\langle &(NN)_{S=0} J/\psi ; J = 1 \ \Big| \ v_{J/\psi-N} \ \Big| 
          (NN)_{S=0} J/\psi ; J = 1 \Big\rangle \nonumber \\
      &=  \frac{1}{3} \ \Big\langle (NJ/\psi)_{S=1/2} N ;  J = 1 \ \Big| \ 
          v_{J/\psi-N} \ \Big| (NJ/\psi)_{S=1/2} N ;  J = 1 \Big\rangle   \nonumber  \\ 
     & \qquad + \frac{2}{3} \ \Big\langle (NJ/\psi)_{S=3/2} N ;  J = 1 \ \Big| \ 
          v_{J/\psi-N} \ \Big| (NJ/\psi)_{S=3/2} N ;  J = 1 \Big\rangle   \nonumber  \\ 
     &= \frac{1}{3} \ \Big[ (v_{0} - v_{s})e^{-\mu r^{2}} \Big]
       + \frac{2}{3} \ \Big[ (v_{0} + \frac{1}{2}v_{s})e^{-\mu r^{2}} \Big]  \nonumber  \\ 
     &=  \ v_{0} e^{-\mu r^{2} } \ .
\end{alignat}
Thus
\begin{alignat}{1}
  V_{\rm eff}^{(1,1)} = \ v_{0} \ .
\end{alignat}
\item $J=0$, $S_{NN}=1$ ($T=0$)

For $J=0$, the only possible state is $S_{NN}=1$ and $S_{J/\psi - N}=1/2$, 
\begin{equation}
  \Big| (NN)_{S=1} J/\psi ; J = 0 \Big\rangle 
     = \Big| (NJ/\psi)_{S=1/2} N ; J = 0 \Big\rangle .
\end{equation}
We obtain 
\begin{alignat}{5}
  \Big\langle &(NN)_{S=1} J/\psi ; J = 0 \ \Big| \ v_{J/\psi-N} \ \Big| 
      (NN)_{S=1} J/\psi ; J = 0 \Big\rangle    \nonumber \\
      &= \ \Big\langle (NJ/\psi)_{S=1/2} N ;  J = 0 \ \Big| 
          v_{J/\psi-N} \ \Big| (NJ/\psi)_{S=1/2} N ;  J = 0 \Big\rangle  \nonumber  \\ 
     &= \ (v_{0} - v_{s})e^{-\mu r^{2}} 
\end{alignat}
and
\begin{alignat}{1}
  V_{\rm eff}^{(0,0)} = \ v_{0} - v_{s} = v_{\rm eff}(1/2) \ .
\end{alignat}

\item $J=2$, $S_{NN}=1$ ($T=0$)

Similarly, for $J=2$, we have $S_{NN}=1$ and $S_{J/\psi - N}=3/2$ \ ,
\begin{equation}
   \Big| (NN)_{S=1} J/\psi ; J = 2 \Big\rangle 
      = \Big| (NJ/\psi)_{S=3/2} N ;  J = 2 \Big\rangle     
\end{equation}
and then
\begin{alignat}{5}
  \Big\langle &(NN)_{S=1} J/\psi ; J = 2 \ \Big| \ v_{J/\psi-N} \ \Big| 
       (NN)_{S=1} J/\psi ; J = 2 \Big\rangle    \nonumber \\
    &= \ \Big\langle (NJ/\psi)_{S=3/2} N ;  J = 2 \ \Big| \ 
          v_{J/\psi-N} \ \Big| (NJ/\psi)_{S=3/2} N ;  J = 2 \Big\rangle   \nonumber  \\ 
    &= \ (v_{0} + \frac{1}{2}v_{s})e^{-\mu r^{2}} \ .
\end{alignat}
We obtain 
\begin{alignat}{1}
  V_{\rm eff}^{(2,0)} = \ v_{0} + \frac{1}{2} v_{s} = v_{\rm eff}(3/2) \ .
\end{alignat}
\end{enumerate}



%



\end{document}